\documentclass[11pt]{article}
\usepackage[margin=1in]{geometry}
\usepackage{amsmath,amssymb,amsthm}
\usepackage{graphicx}
\usepackage{natbib}
\usepackage{hyperref}
\usepackage{booktabs}

\newtheorem{proposition}{Proposition}
\newtheorem{corollary}{Corollary}

\theoremstyle{definition}
\newtheorem{definition}{Definition}

\title{The Winner's Bliss in Common-Value Auctions under Horizontal Differentiation}
\author{Jiawei Chen\\ UC Irvine \and Anh Nguyen\\ Carnegie Mellon \and Matthew Shum\\ Caltech}
\date{\today}

\begin{document}
\maketitle

\begin{abstract}
We study common-value auctions in which bidders have horizontally differentiated preferences. In a specific two-bidder parameterization, winning conveys good news about the object's value to the winner, a phenomenon we call the ``winner's bliss,'' in contrast to the conventional winner's curse. Additional implications also differ from the conventional analysis.  When bidders' preferences are horizontally differentiated, seller revenue is decreasing in information disclosure, and advantageous selection sustains bilateral trade under asymmetric information.
\end{abstract}

\smallskip\noindent\textbf{Keywords:} Winner's curse, auctions, common values, horizontal differentiation, advantageous selection.\\
\textbf{JEL Codes:} D44, D82, G14, L86.

\section{Introduction}

The winner's curse is among the most studied phenomena in auction theory. In a common-value auction, winning reveals that opponents had less optimistic signals, causing the winner to revise downward her valuation \citep{thaler1988, milgrom1982}. Recent work identifies circumstances where the opposite holds. \citet{bergemann2020} show that revenue-optimal mechanisms can produce a ``winner's blessing'' by skewing allocation toward lower-signal bidders. \citet{lauermann2023} find a related effect when the number of competitors is uncertain. Both channels require specific features of the environment, namely seller optimization or market-thickness uncertainty.

We identify a different channel, namely \emph{horizontal preference heterogeneity}. When bidders disagree on which state makes the object valuable, winning reveals that the state is favorable to the winner. We call this the ``winner's bliss.'' It arises in the equilibrium of a standard first-price auction without mechanism design or uncertainty about the number of bidders.

The winner's bliss intuition is straightforward in the context of online advertising.\footnote{See \citet{alcobendas2021} for empirical evidence and discussions on how cookies in online ad auctions communicate noisy information on customer characteristics.} A cosmetics brand targets women; a beer brand targets men. Each observes a noisy signal about an internet user's unknown gender (the state). If the cosmetics brand wins, she infers that the beer brand also received a signal indicating the user is female, confirming rather than contradicting her own signal.

We introduce a simple two-bidder model permitting equilibrium characterization.  In this model, bidder preferences are parameterized  by $\mu \in [0,1]$, the probability that the two bidders share the same type, with $\mu=0$ denoting horizontally differentiated preferences, and $\mu=1$ indicating horizontal alignment. Our main results are:
\begin{enumerate}
\item  Winner's bliss holds with horizontally differentiated preferences ($\mu < 1/2$); the conventional winner's curse holds with aligned preferences ($\mu > 1/2$). (Prop.~\ref{prop:bliss})
\item  Revenue is decreasing in preference alignment, and revealing the state reduces revenue under differentiated preferences but raises it when aligned.  This echoes Board's (\citeyear{board2009}) allocation effect  and Ganuza's (\citeyear{ganuza2004}) ``ignorance promotes competition'' result.  (Props.~\ref{prop:revenue} and~\ref{prop:antilinkage})
\item  In bilateral trade, horizontally differentiated preferences produce advantageous selection that sustains exchange (at ~90\% of first-best efficiency in our parameterization).  This contrasts with adverse selection that impedes trade under aligned preferences. (Prop.~\ref{prop:trade})
\end{enumerate}

\paragraph{Related literature.} 

Our work relates to papers showing that private information need not exacerbate the winner's curse or adverse selection. Ganuza and Penalva (2010) show that signal structures affect auction outcomes through the dispersion of bidders' posterior valuations. de Meza and Webb (2001) study advantageous selection in insurance markets, and Fieseler et al. (2003) show that negatively interdependent valuations can facilitate efficient trade. Finally, Loertscher and Muir (2025) study revenue-maximizing mechanisms for horizontally differentiated goods in a private-value Hotelling framework. In contrast, our setting is common-value: bidders receive signals about an unknown state, and horizontal preference heterogeneity makes winning good news about the object's value.

\section{Model}\label{sec:model}
We introduce a simple two-bidder first-price auction model permitting equilibrium and analytical characterization.
An object has unknown state $\theta \in \{A, B\}$ with $\Pr(\theta = A) = 1/2$. Two risk-neutral bidders each have private type $t_i \in \{A, B\}$; bidder $i$ receives utility $u_i=1$ if $\theta = t_i$ and $u_i=0$ otherwise. Types satisfy $\Pr(t_1 = t_2) = \mu \in [0, 1]$, with each type equally likely. Each bidder knows her own type but not her opponent's.

Each bidder observes an independent signal $S_i \sim U[0,1]$ with
\[
\Pr(\theta = A \mid S_i = s_i, S_j = s_j) = \frac{s_i + s_j}{2}.
\]
A type-$A$ bidder's expected value for the object is $v_A(s_i, s_j) \equiv \mathbb{E}[u_i \mid s_i, s_j] = (s_i + s_j)/2$; a type-$B$ bidder's is $v_B(s_i, s_j) \equiv \mathbb{E}[u_i \mid s_i, s_j] = 1 - (s_i + s_j)/2$. The model has common values (each bidder's value depends on both signals) but signals are independent.\footnote{Common-value auctions with independent signals are studied in the wallet game \citep{klemperer1998} and in \citet{hendricks2003}; our model shares this structure.} At $\mu = 1$ this is a standard common-value auction; at $\mu = 0$ it captures pure horizontal differentiation. 

While the parameterization above has been chosen for analytical tractability, the following interpretation may be useful. An airline holds a special vacation auction. The flight's destination (beach vs. ski resort) will be chosen based on passengers' preferences. A beach-lover derives value only from a beach destination; a ski-lover only from a mountain one. Two passengers  bid for a seat, utilizing their signals of their fellow passengers' preferences.  Neither knows the other's preference.

\subsection{Equilibrium and the Winner's Bliss}\label{sec:equilibrium}

We look for a symmetric Bayes-Nash equilibrium in monotone bidding strategies. Let $\beta_A(s)$ and $\beta_B(s)$ denote the equilibrium bid functions for type-$A$ and type-$B$ bidders, respectively. By type-symmetry of the model, $\beta_B(s) = \beta_A(1-s)$. Proofs of all propositions are collected in the Appendix.

\begin{proposition}\label{prop:equilibrium}
There exists a type-symmetric monotone equilibrium:
\begin{equation}\label{eq:betaA}
\beta_A(s) = \frac{1-\mu}{2} + \frac{\mu s}{2}, \qquad \beta_B(s) = \frac{1 - \mu s}{2}.
\end{equation}
\end{proposition}

The monotone equilibrium exists for $\mu \in (0,1]$ but not at $\mu = 0$. When the opponent is known to have opposite preferences, a bidder with the lowest signal can profitably deviate upward, breaking monotonicity. This reflects the tractability of our parameterization; in more general settings with horizontal preferences, monotone equilibria need not exist.

\begin{definition}\label{def:bliss}
\emph{Winner's bliss} occurs at signal $s$ if $\mathbb{E}[v_A \mid s, \text{win}] > \mathbb{E}[v_A \mid s]$; \emph{winner's curse} if the inequality reverses.
\end{definition}

\begin{proposition}\label{prop:bliss}
In the equilibrium of Proposition~\ref{prop:equilibrium},
\[
\mathbb{E}[v_A \mid s, \text{win}] - \mathbb{E}[v_A \mid s] = \frac{(2\mu - 1)(s-1)}{4}.
\]
Winner's bliss holds for all $s \in [0,1)$ when $\mu < 1/2$; winner's curse when $\mu > 1/2$.
\end{proposition}

\begin{figure}[htbp]
\centering
\includegraphics[width=\textwidth]{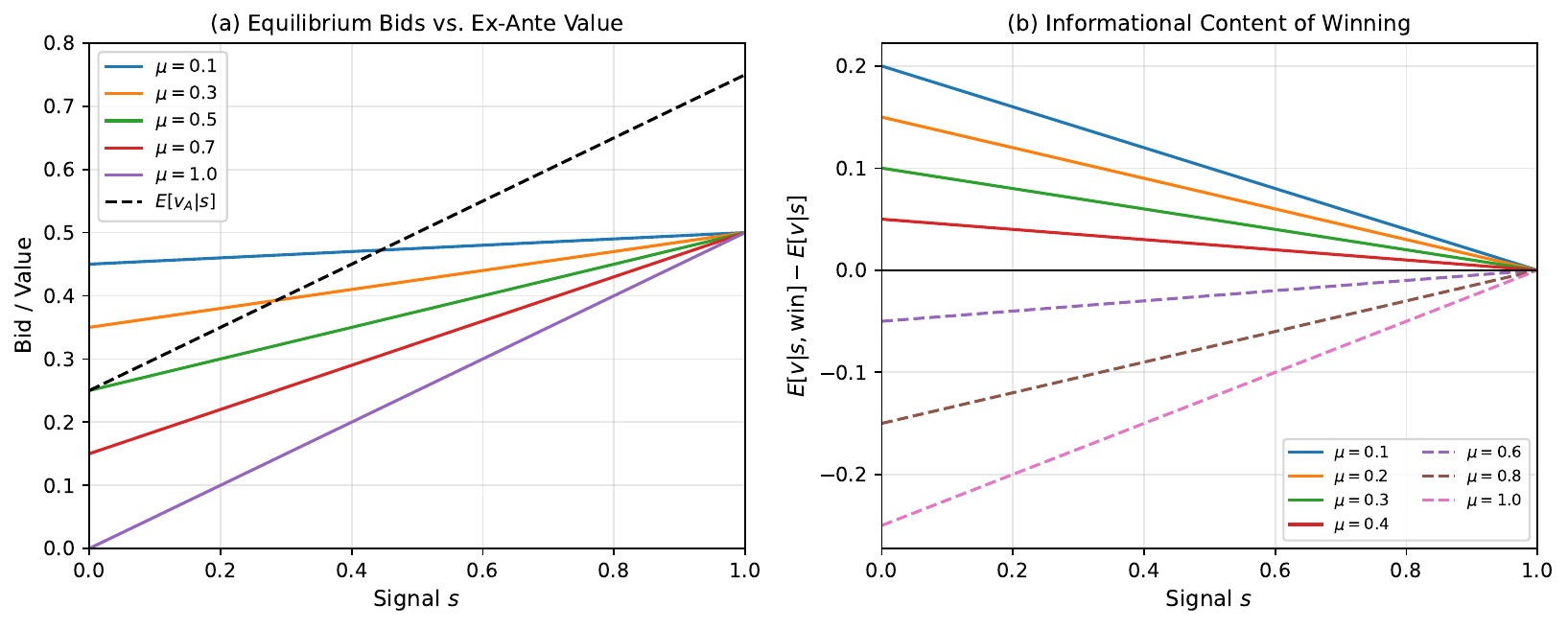}
\caption{\small (a) Equilibrium bidding strategies for different $\mu$. Dashed line: ex-ante value $\mathbb{E}[v_A \mid s]$. For $\mu < 0.5$, bids exceed ex-ante value at low signals. (b) Informational content of winning: positive values indicate winner's bliss ($\mu < 0.5$), negative values indicate winner's curse ($\mu > 0.5$).}
\label{fig:bliss}
\end{figure}

The threshold $\mu = 1/2$ is where the opposing-type channel (winning confirms the bidder's signal) exactly offsets the same-type channel (winning means weaker opponent signal). The effect is largest at low signals and vanishes as $s$ approaches 1. Figure~\ref{fig:bliss} illustrates this pattern.

\begin{proposition}\label{prop:revenue}
Expected revenue is $R(\mu) = 1/2 - \mu/6$, strictly decreasing in $\mu$.
\end{proposition}

Revenue is maximized at $\mu = 0$ (opposing preferences) and minimized at $\mu = 1$ (aligned preferences).

\subsection{Information Disclosure under Winner's Bliss}\label{sec:antilinkage}

In some settings such as online ad auctions, platforms control how much user information they disclose to the competing bidders (advertisers). With horizontally differentiated bidders, disclosure can reduce revenue.

To explore this, consider a platform that reveals the state $\theta$
publicly with probability $q$ and conceals it otherwise. Under concealment, bidders
rely on their private signals and revenue is $1/2 - \mu/6$ by
Proposition~\ref{prop:revenue}. Under revelation, $\theta$ becomes common knowledge and
the bidders compete in a private-value auction.

\begin{proposition}[Allocation Effect]\label{prop:antilinkage}
Under full revelation of the state, expected revenue is $\mu/2$. Hence under
disclosure probability $q$,
\[
R(q) = q\,\frac{\mu}{2} + (1-q)\left(\frac{1}{2} - \frac{\mu}{6}\right),
\]
which is decreasing in $q$ when $\mu < 3/4$ and increasing when $\mu > 3/4$.
\end{proposition}

\begin{figure}[htbp]
\centering
\includegraphics[width=\textwidth]{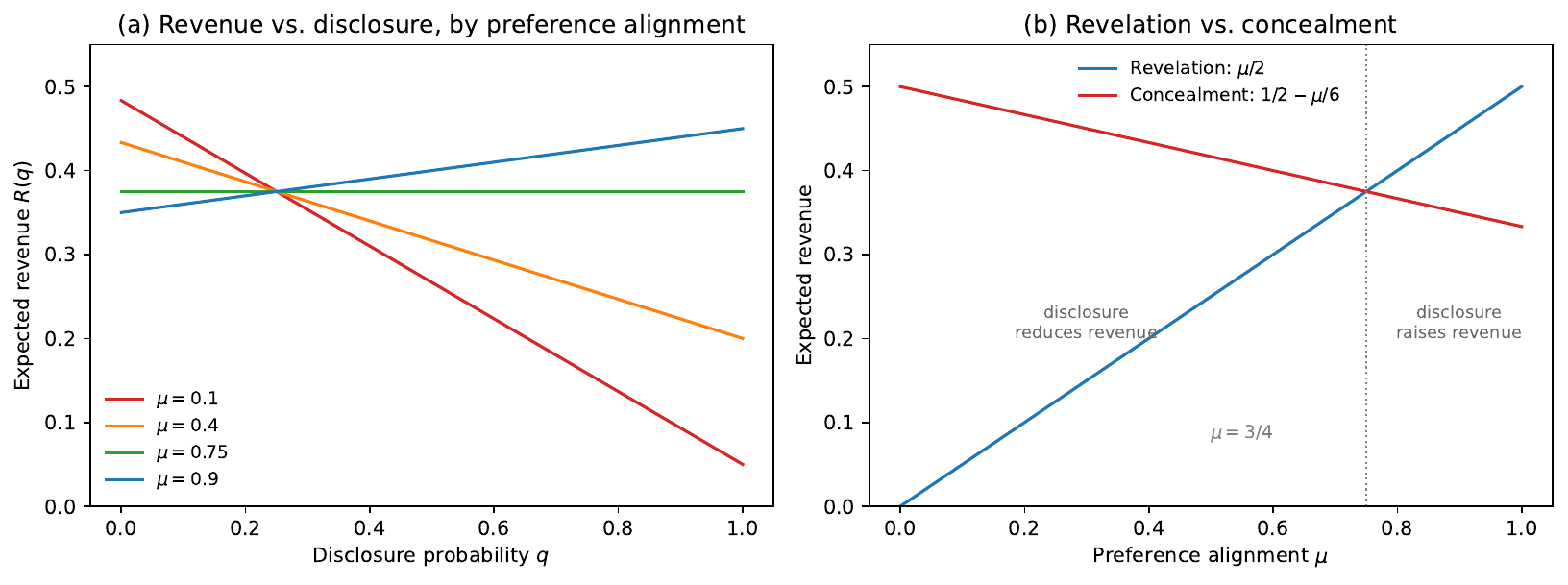}
\caption{\small (a) Expected revenue $R(q)$ versus the disclosure probability $q$ for
several levels of preference alignment $\mu$: decreasing when $\mu < 3/4$ and
increasing when $\mu > 3/4$. (b) Revelation revenue $\mu/2$ versus concealment
revenue $1/2 - \mu/6$; the two cross at $\mu = 3/4$, so disclosure reduces revenue
for opposed preferences and raises it for aligned ones.}
\label{fig:antilinkage}
\end{figure}

Revealing the state thus reduces revenue when preferences are sufficiently opposed
($\mu < 3/4$) and raises it when sufficiently aligned ($\mu > 3/4$). This is the
allocation effect of \citet{board2009}, whereby revealing which bidder values the object can
soften competition. The linkage principle, under which disclosure raises
revenue, holds only in the aligned region; under opposing preferences the allocation
effect dominates. This threshold need not coincide with the selection threshold $\mu = 1/2$ of Proposition~\ref{prop:bliss}, which marks where winning switches from good to bad news. Figure~\ref{fig:antilinkage} illustrates this result.


\subsection{Advantageous Selection in Bilateral Trade}\label{sec:trade}

We apply the model to exchange. Trader~1 (buyer, type~$A$) values the asset at $(s_1 + s_2)/2$; Trader~2 (seller, type~$B$) values it at $1 - (s_1 + s_2)/2$. Trade occurs at exogenous price $p$.

This setting has genuine gains from trade only when $s_1 + s_2 > 1$ (i.e., $v_A > v_B$), placing it outside the \citet{milgrom1982stokey} no-trade premise (which requires ex ante efficiency of the initial allocation). The question is whether information asymmetry facilitates or impedes trade. Under aligned preferences, the counterparty's willingness to trade is bad news (adverse selection). Under horizontal preferences, it is good news (advantageous selection). A related observation appears in \citet{fieseler2003}, who show that efficient mechanisms are easier to construct when valuations are negatively interdependent.

Consider threshold strategies where the buyer purchases if $s_1 > a^*$, while the seller sells if $s_2 > b^*$.
\begin{proposition}\label{prop:trade}
At price $p = 1/2$, the unique equilibrium in threshold strategies  is $a^* = b^* = 1/3$:
\begin{itemize}
\item Trade probability: $4/9 \approx 44.4\%$. First-best: $1/2$.
\item Realized surplus efficiency: $8/9 \approx 88.9\%$.
\end{itemize}
\end{proposition}

\begin{corollary}[Advantageous Selection]\label{cor:advantageous}
Both parties revise valuations favorably upon trade:
$\mathbb{E}[v_A \mid s_1, S_2 > 1/3] > \mathbb{E}[v_A \mid s_1]$ and $\mathbb{E}[v_B \mid s_2, S_1 > 1/3] < \mathbb{E}[v_B \mid s_2]$.
The counterparty's willingness to trade is good news, producing a positive feedback loop that sustains exchange.
\end{corollary}

\begin{figure}[htbp]
\centering
\includegraphics[width=\textwidth]{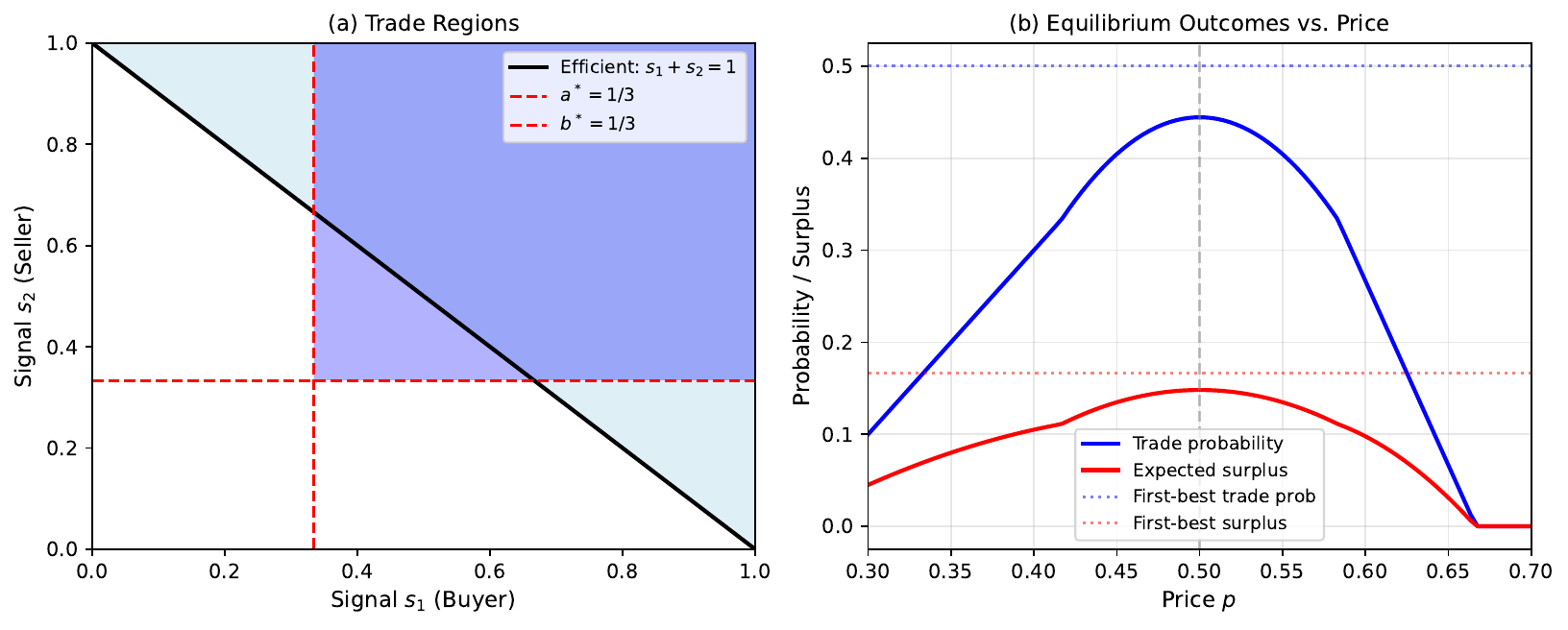}
\caption{\small (a) Trade regions in $(s_1, s_2)$ space. Dark region: equilibrium trade ($s_1 > 1/3$ and $s_2 > 1/3$); light region: efficient trade ($s_1 + s_2 > 1$). (b) Trade probability and surplus as functions of price $p$; the symmetric price $p = 1/2$ maximizes surplus.}
\label{fig:trade}
\end{figure}

As an application, in financial markets with heterogeneous investor styles (value vs.\ growth, domestic vs.\ foreign), this predicts that style diversity facilitates trade.  Figure~\ref{fig:trade} displays the equilibrium trade regions and surplus.

\section{Conclusion}\label{sec:conclusion}

Horizontal preference heterogeneity produces a winner's bliss in common-value auctions.\footnote{A related logic applies in committee voting \citep{feddersen1998}. Agreement from an ideologically opposed member is more informative than agreement from a like-minded one, since the opposed member must have received a strong signal to overcome her lean.} The resulting sign reversal in selection propagates to information disclosure and bilateral trade, with testable implications for online advertising markets. Companion experimental work \citep{nguyen2023} provides initial laboratory evidence. The same logic may extend to two-sided matching markets under incomplete information, where being chosen by a participant with different preferences is more informative about match quality than being chosen by a similar one.

\bibliographystyle{aer}

\appendix
\section*{Appendix: Proofs}

\paragraph{Proof of Proposition~\ref{prop:equilibrium}.}
A type-$A$ bidder with signal $s$ who bids $b$ wins against a same-type opponent (probability $\mu$) when that opponent's signal $s' < \beta_A^{-1}(b)$, and against an opposing-type opponent (probability $1-\mu$) when $s' > 1 - \beta_A^{-1}(b)$. Her expected profit is
\[
\Pi(b, s) = \mu \int_0^{\beta_A^{-1}(b)} \!\left[\frac{s+s'}{2} - b\right] ds' \;+\; (1-\mu) \int_{1-\beta_A^{-1}(b)}^1 \!\left[\frac{s+s'}{2} - b\right] ds'.
\]
Setting $\partial\Pi/\partial b = 0$ using $d\beta_A^{-1}/db = 1/\beta_A'$ and evaluating at $b = \beta_A(s)$ (so that $\beta_A^{-1}(b) = s$) yields
\[
\mu\!\left(\frac{s - \beta_A(s)}{\beta_A'(s)}\right) + (1-\mu)\!\left(\frac{1/2 - \beta_A(s)}{\beta_A'(s)}\right) = s,
\]
which simplifies to $\big[\mu s + (1-\mu)/2 - \beta_A(s)\big] = \beta_A'(s)\,s$, i.e., $\frac{d}{ds}\!\big[s\,\beta_A(s)\big] = \mu s + (1-\mu)/2$. Integrating:
\[
s\,\beta_A(s) = \frac{\mu s^2}{2} + \frac{(1-\mu)s}{2} + C.
\]
Boundedness as $s \to 0$ requires $C = 0$, giving $\beta_A(s) = (1-\mu)/2 + \mu s/2$. For global optimality, a bidder with signal $s$ who deviates to $\beta_A(s')$ earns payoff $U(s', s)$ with $\partial U/\partial s' \propto (s - s')$, which is positive for $s' < s$ and negative for $s' > s$, confirming $s' = s$ is optimal.\footnote{In companion experimental work \citep{nguyen2023}, subjects' bid sensitivity to signals increases with $\mu$, consistent with the equilibrium prediction.} \hfill$\square$

\paragraph{Proof of Proposition~\ref{prop:bliss}.}
In equilibrium, a type-$A$ bidder wins against a same-type opponent when $s_j < s$ (probability $s$) and against an opposing-type opponent when $s_j > 1-s$ (probability $s$). Thus $\Pr(\text{win} \mid s) = s$. The conditional expected value:
\begin{align*}
\mathbb{E}[v_A \mid s, \text{win}] &= \frac{\mu \cdot s \cdot \tfrac{3s}{4} + (1-\mu) \cdot s \cdot \tfrac{s+2}{4}}{s} = \frac{s(2\mu+1)}{4} + \frac{1-\mu}{2}.
\end{align*}
Since $\mathbb{E}[v_A \mid s] = s/2 + 1/4$, subtraction gives the result. \hfill$\square$

\paragraph{Proof of Proposition~\ref{prop:revenue}.}
Revenue equals the expected winning bid. With probability $\mu/2$ both bidders are type~$A$; the winner has the higher signal $s_{(2:2)}$ with density $2s$, paying $\beta_A(s_{(2:2)})$. With probability $\mu/2$ both are type~$B$ (winner has the lower signal). With probability $1-\mu$ types differ; type~$A$ wins when $s_A + s_B > 1$. Integrating across cases yields $R = 1/2 - \mu/6$. \hfill$\square$

\paragraph{Proof of Proposition~\ref{prop:antilinkage}.}
Condition on $\theta = A$. A type-$A$ bidder has value~$1$ and a type-$B$ bidder value~$0$; the latter bids~$0$. A value-$1$ bidder faces a rival who shares her value with probability $\mu$ and uses the symmetric mixed strategy with c.d.f.\ $G(b) = \frac{1-\mu}{\mu}\frac{b}{1-b}$ on $[0,\mu]$, equating her payoff $[(1-\mu) + \mu G(b)](1-b)$ to $1-\mu$ over the support. Given $\theta = A$, both bidders are type~$A$ with probability $\mu/2$, exactly one is type~$A$ with probability $1-\mu$, and neither is with probability $\mu/2$; computing the expected higher bid across these cases gives $\mu/2$. Combining with the concealment revenue from Proposition~\ref{prop:revenue} gives $R(q)$, and $\partial R/\partial q = 2\mu/3 - 1/2$. \hfill$\square$

\paragraph{Proof of Proposition~\ref{prop:trade}.}
Buyer purchases if $\mathbb{E}[v_A \mid s_1, S_2 > b^*] \geq p$, giving $a^* = 2p - (1+b^*)/2$. Seller sells if $p \geq \mathbb{E}[v_B \mid s_2, S_1 > a^*]$, giving $b^* = 2(1-p) - (1+a^*)/2$. At $p=1/2$: $a^* = (1-b^*)/2$, $b^* = (1-a^*)/2$, yielding $a^* = b^* = 1/3$. Expected surplus is $\int_{1/3}^{1}\int_{1/3}^{1}(s_1+s_2-1)\,ds_2\,ds_1 = 4/27$; first-best surplus is $\int\int_{\{s_1+s_2>1\}}(s_1+s_2-1)\,ds = 1/6$. The ratio is $(4/27)/(1/6) = 8/9$. \hfill$\square$

\end{document}